\documentclass[12pt]{article}

\usepackage{amsmath,amssymb,amsfonts,amsbsy}
\usepackage{cite}
\usepackage{graphicx}
\usepackage{wrapfig}
\def\xb{\overline{x}}

\def\als{\alpha_s}
\def\vk{{\bf k}_{\perp}}
\def\vbs{{\bf b}}

\begin{document}
\addcontentsline{toc}{subsection}{{Cross sections and spin
asymmetries in vector meson leptoproduction.}\\
{\it S.V. Goloskokov}}

%%%%%%% please do not touch these! %%%%%%
\setcounter{section}{0}
\setcounter{subsection}{0}
\setcounter{equation}{0}
\setcounter{figure}{0}
\setcounter{footnote}{0}
\setcounter{table}{0}

\begin{center}
\textbf{Generelized Parton Distributions in light meson production.}

\vspace{5mm}

S.V. Goloskokov\footnote{Presented at the 5th joint International
HADRON STRUCTURE '11 Conference, June 27- July 1, 2011, Tatranska'
Strba (Slovak Republic)}

\vspace{5mm}

\begin{small}
Bogoliubov Laboratory of Theoretical Physics, Joint Institute for
Nuclear Research, Dubna 141980, Moscow region, Russia
\end{small}
\end{center}

\vspace{0.0mm} % Don't laugh: it does change the spacing!

\begin{abstract}
We analyze light  meson electroproduction  within the handbag
model, where the amplitude factorizes into Generelized Parton
Distributions (GPDs) and a hard scattering part. The cross
sections and spin asymmetries for various vector and pseudoscalar
mesons are analyzed. We discuss what information on hadron
structure can be obtained from GPDs.
\end{abstract}

\vspace{7.2mm}

 \section{Introduction}
 In this report, we study leptoproduction of light mesons at  small
momentum transfer and large photon virtualities $Q^2$ on the basis
of the handbag approach.  In this kinematic region the
 leading twist amplitude factorizes into a hard meson
electroproduction subprocess off partons and  GPDs \cite{fact}.
The hard subprocess is calculated within the modified perturbative
approach (MPA)   where quark transverse degrees of freedom
accompanied by Sudakov suppressions are considered. The quark
transverse momentum regularizes the end-point singularities in the
higher twist amplitudes so that it can be calculated in the model.
The GPDs which contains information on the soft physics are
constructed using double distributions. They depend on  $\bar x$
-the momentum fraction of the parton, $\xi$- skewness which is
determined as a difference of the parton momenta and related to
Bjorken- $x_B$ as $\xi \sim x_B/2$ and $t$-momentum transfer.

Within our approach \cite{gk05,gk06,gk07q} we calculate the gluon
and valence quark GPDs $H$ contribution  to the meson production
amplitudes and, subsequently, the cross sections and the spin
observables in the light meson electroproduction off unpolarized
proton target. Using GPDs $E$ we extend our analysis to a
transversally polarized target. We calculate cross sections and
the $A_{UT}$ asymmetry for various vector meson productions. Our
results \cite{gk05,gk06,gk07q,gk08} on meson electroproducion are
in good agreement with experimental data in the  HERA
\cite{h1,zeus} COMPASS \cite{compass} and HERMES \cite{hermes}
energy range.

The study of pseudoscalar meson electroproduction \cite{gk09,gk11}
gives access to the GPD $\tilde H$ and $\tilde E$. In the
description of this reaction the twist-3 effects are very
essential too. Within the handbag approach these twist-3 effects
were modeled by the transversity GPDs, in particular, $H_T$ and
$E_T$ in conjunction with the twist-3 pion wave function. Our
results \cite{gk09,gk11} on the cross section and moments of spin
asymmetries for the polarized target are in good agreement with
HERMES experimental data.

 \section{GPDs and mesons production amplitudes in the handbag approach}

In the handbag model, the amplitude of the vector meson production
off the proton with positive helicity reads as a convolution of
the hard partonic subprocess
 ${\cal H}^a$ and GPDs $H^a\,(\widetilde{H}^a)$

\begin{equation}\label{amptt}
{\cal M}^{a}_{\mu'\pm,\mu +} = \, \sum_{a}\,[ \langle {H}^a
  \rangle+O(\langle \widetilde{H}^a
  \rangle ) ]
\end{equation}

$$\langle {H}^a\rangle = \sum_{\lambda}
         \int_{xi}^1 d\xb
        {\cal H}^{a}_{\mu'\lambda,\mu \lambda}(Q^2,\xb,\xi,t)
                                   \hat H^{a}(\xb,\xi,t),$$
where  $a$ denotes the gluon and quark contribution with the
corresponding flavors;
 $\mu$ ($\mu'$) is the helicity of the photon (meson), and $\xb$
 is the momentum fraction of the
parton with helicity $\lambda$.  In the region of small $\bar x
\leq 0.01$  gluons give the dominant contribution. At larger $\bar
x \sim 0.1$  the   quark contribution plays an important role
\cite{gk06}.

The subprocess amplitude is calculated within the MPA
\cite{sterman}. The  amplitude ${\cal H}^a$ is   a contraction of
the hard part ${\cal F}^a$, which is calculated perturbatively and
includes the transverse quark momentum $\vk$, and the
nonperturbative $\vk$-dependent meson wave function
\cite{koerner}. The gluonic corrections are treated in the form of
the Sudakov factors in the hard partonic subprocess ${\cal H}^a$.
The resummation and exponentiation of the Sudakov corrections can
be done  in the impact parameter space \cite{sterman}. We use the
Fourier transformation to transfer integrals from the $\vk$ to
$\vbs$ space. Within the MPA the subprocess amplitude in the
impact parameter (${\bf b}$) space reads
\begin{eqnarray}\label{aa}
{\cal H}^a_{\mu'\lambda,\mu\lambda} = \int d\tau d^2b\,
         \hat{\Psi}(\tau,-\vbs,\mu_F)\,
      \hat{\cal F}^{a}_{\mu'\lambda,\mu\lambda}(\xb,\xi,\tau,Q^2,\vbs,\mu_R)\,
      \als(\mu_R)\,\nonumber\\
        \times{\rm
      exp}{[-S(\tau,\vbs,Q^2,\mu_F,\mu_R)]}.
\end{eqnarray}
The Sudakov factor $S$ and the choice of the renormalization
($\mu_R$) and factorization ($\mu_F$) scales can be found in
\cite{gk05,gk06}. The hard scattering kernels ${\cal F}$ and
$\hat{\Psi} (\tau,-\vbs)$ in (\ref{aa}) are the Fourier transform
of the $\vk$ dependent subprocess amplitude, and the meson wave
function ($\tau$ ($1-\tau$) is the momentum fraction of the quark
(antiquark) that enters into the meson, defined with respect to
the meson momentum).

The $\hat H^{a}$ in (\ref{amptt}) is expressed in terms of GPDs
which contain the extensive information on the hadron structure.
At zero skewness and momentum transfer GPD becomes equal to the
corresponding parton distribution function(PDF). The form factors
of hadrons are the first moments of the corresponding GPDs,
\begin{eqnarray}\label{ff}
\int dx\, H^a(x,\xi,t)= F^a_1(t);\nonumber\\
\int dx\,E^a(x,\xi,t)= F^a_2(t),
\end{eqnarray}
 where  $F^a_1$ and $F^a_2$ are the
Dirac and Pauli form factors with flavor $a$. Information on the
parton angular momenta can be obtained from the Ji sum rules
\cite{ji}
\begin{equation}\label{ji}
  J^q =\frac{1}{2}\int x dx (H^q(x,\xi,0)+E^q(x,\xi,0)).
\end{equation}

To estimate  GPDs, we use the double distribution (DD)
representation \cite{mus99}
\begin{eqnarray}\label{ddr}
  H_i(\xb,\xi,t) =  \int_{-1}
     ^{1}\, d\beta \int_{-1+|\beta|}
     ^{1-|\beta|}\, d\alpha \delta(\beta+ \xi \, \alpha - \xb)\, f_i(\beta,\alpha,t)
\end{eqnarray}
which connects  GPDs with PDFs through the DD function $f$,
\begin{eqnarray}\label{ddf}
f_i(\beta,\alpha,t)= h_i(\beta,t)\,
                   \frac{\Gamma(2n_i+2)}{2^{2n_i+1}\,\Gamma^2(n_i+1)}\,
                   \frac{[(1-|\beta|)^2-\alpha^2]^{n_i}}
                           {(1-|\beta|)^{2n_i+1}}.
\end{eqnarray}
The functions $h_i$  are expressed in terms of PDFs and
parameterized as
\begin{equation}\label{pdfpar}
h(\beta,t)= N\,e^{b_0 t}\beta^{-\alpha(t)}\,(1-\beta)^{n}.
\end{equation}
Here  the $t$- dependence is considered in a Regge form and
$\alpha(t)$ is a corresponding Regge trajectory.

 From different meson productions at moderate HERMES and COMPASS
 energies we can get information about valence
and sea quark effects. Really, the quarks contribute to meson
production processes in different combinations. For uncharged
meson production we have standard GPDs and find
\begin{equation}\label{quarks}
\rho:\;\; \propto \frac{2}{3} H^u +\frac{1}{3} H^d;\;\;\omega:\;\;
\propto \frac{2}{3} H^u -\frac{1}{3} H^d.
\end{equation}
For production of charged and strange mesons we have transition $
p\to n$ and $p\to \Sigma$ GPDs and we have
\begin{equation}\label{quarkspl}
\rho^+:\;\; \propto H^u -H^d;\;\;\;\;K^{*0}:\;\;  \propto  H^d
-H^s.
\end{equation}
For pseudoscalar mesons the polarized GPDs contribute as
\begin{equation}
 \pi^+:\;\;\propto \tilde H^u -
\tilde H^d ;\;\;\pi^0:\;\; \propto \frac{2}{3} \tilde H^u
+\frac{1}{3} \tilde H^d.
\end{equation}
The same combinations are valid for $E, \tilde E$ contributions.
Thus, we can test various GPDs  in the mentioned reactions.

\section{Vector meson electroproduction}
In this section, we  study vector meson electroproduction and
consider the cross section and spin observables using amplitudes
and GPDs calculated in the previous section. The $H$ GPDs are
modeled by  DD (\ref{ddr}),(\ref{ddf}). The parameters in
(\ref{pdfpar}) are determined from the CTEQ6 analysis of PDFs
\cite{CTEQ}. More details together with parameters of the model
can be found in \cite{gk05,gk06,gk07q}.

 Our approach has been used to analyze
data on $\rho^0$ and $\phi$ electroproduction in a wide energy
range. We consider the gluon, sea and quark GPD $H$ contribution
to the meson production amplitudes. This permits us to study
vector meson production from moderate HERMES energies $W \sim 5
\mbox{GeV}$ till high HERA energies $W \sim 75 \mbox{GeV}$
\cite{gk06,gk07q}. The $k_\perp^2/Q^2$ corrections in propagators
of the hard subprocess amplitudes are extremely important at low
$Q^2$ and permit us to describe experimental data properly. If we
omit $k_\perp^2/Q^2$ corrections (leading twist approximation),
the cross section increases by a factor of about 10 at $Q^2 \sim
3\mbox{GeV}^2$ (see Fig. 1).
\begin{figure}[h!]
\centerline{
\includegraphics[width=0.6\textwidth]{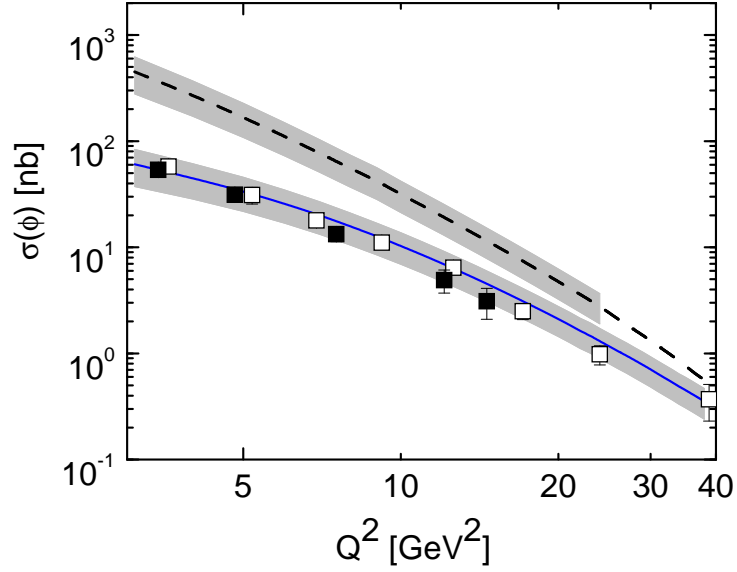}}
\caption{Longitudinal cross sections of $\phi$ production at $W=75
\mbox{GeV}$. Data are from H1  -solid symbols and ZEUS -open
symbols. Dashed line - leading twist result.}
\end{figure}

The obtained results \cite{gk06,gk07q}  are in good agreement with
the experiments at HERA \cite{h1,zeus}, COMPASS \cite{compass},
HERMES \cite{hermes} and E665 \cite{e665} energies for
electroproduced $\rho$ and $\phi$ mesons. If we extend our
analysis to lower energies $W \sim 2.5 \mbox{GeV}$, we find a good
description of the $\phi$ cross section  at CLAS \cite{clas}, as
shown on Fig.~2. This means that we have good  results for gluon
and sea quark contributions which are important in the $\phi$
meson production.
\begin{figure}[h!]
\centerline{
\includegraphics[width=0.6\textwidth]{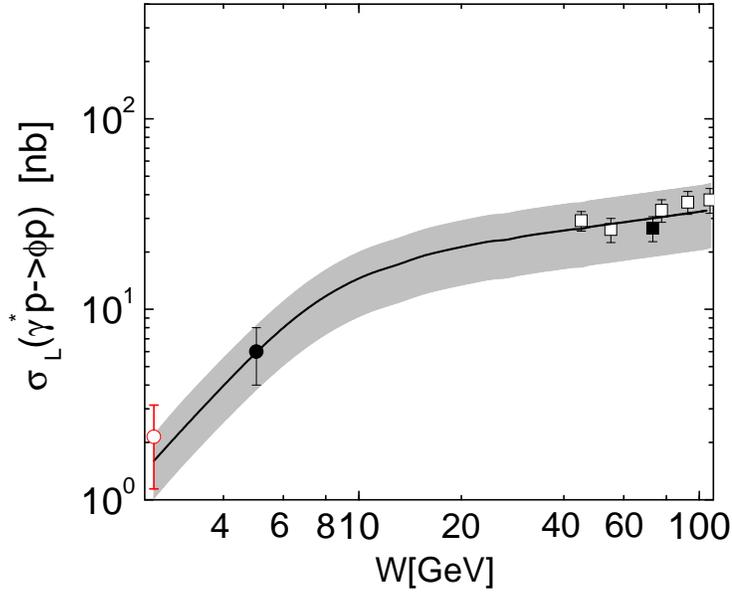}}
\caption{The longitudinal cross section for $\phi$ at
$Q^2=3.8\,\mbox{GeV}^2$. Data: HERMES  (solid circle), ZEUS (open
square), H1 (solid square), open circle-  CLAS data point}
\end{figure}

For the $\rho$ production we find an essential contribution of the
valence quarks which are of the order of gluon and sea effects at
$W \sim 5 \mbox{GeV}$. At lower energies, the quark contribution
decreases, as well as the gluon and sea one. As a result, the
cross section falls at energies $W \le 5\mbox{GeV}$. This is in
contradiction with CLAS \cite{clas} results where  the rapid
growth of $\sigma_\rho$ is observed in this energy range, Fig.~3.
Thus, most probably we have a problem with the valence quark
contribution at low JLAB energies, which is not solved till now.
\begin{figure}[h!]
\centerline{
\includegraphics[width=0.6\textwidth]{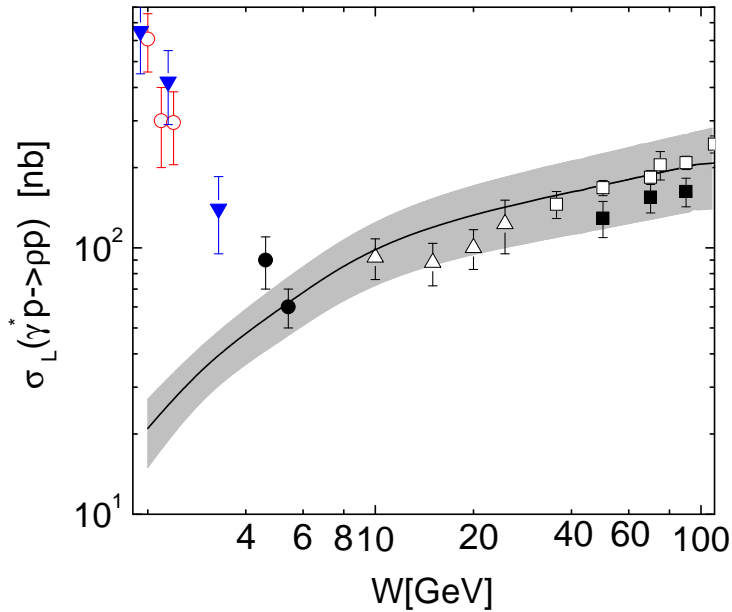}}
\caption{The longitudinal cross section for $\rho$ at
$Q^2=4.0\,\mbox{GeV}^2$. Data: HERMES (solid circle), ZEUS (open
square), H1 (solid square), E665 (open triangle), open circles-
CLAS,  CORNEL -solid triangle}
\end{figure}

Using the quarks contribution (\ref{quarks}), (\ref{quarkspl}) to
meson  amplitudes we can calculate the cross section for different
meson production. In Fig.~4, our results for COMPASS energy $W=10
\mbox{GeV}$ are shown.
\begin{figure}[h!]
\centerline{
\includegraphics[width=0.6\textwidth]{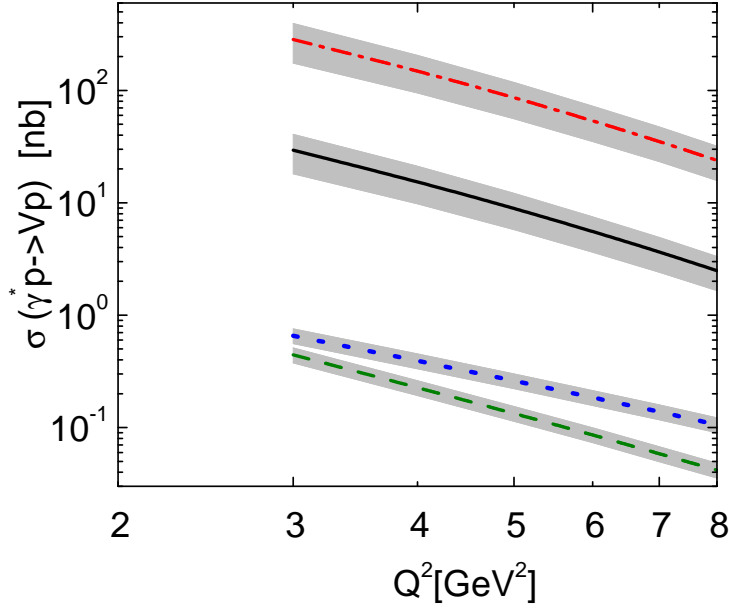}}
\caption{Predicted integrated over $t$ cross section
  at COMPASS $W = 10\mbox{GeV}$ energies for various mesons.
    Dotted-dashed line $\rho^0$;
    full line $\omega$; dotted line $\rho^+$ and dashed line $K^{* 0}$.}
\end{figure}

The proton spin-flip amplitude is associated with $E$ GPDs
\begin{eqnarray}
{\cal M}_{\mu' -,\mu +} \propto \frac{\sqrt{-t}}{2 m}
                          \int_{-1}^1 d\xb\,
           E^a(\xb,\xi,t)\, F^a_{\mu',\mu}(\xb,\xi).
\end{eqnarray}

The GPD $E$ is not well known till now. The standard relation
\begin{equation}\label{e}
E^a(x,0,0)=e^a(x)
\end{equation}
takes place where $e^a$ is the corresponding PDF. We constrained
$e^a$  by the Pauli form factors of the nucleon \cite{pauli},
positivity bounds and sum rules. The GPD $E$ is constructed using
DD.

We would like to note that the first moment of $e$ is proportional
to the quark anomalous magnetic moment
\begin{equation}
\int^1_0 dx e^a_{val}(x)=\kappa^a.
\end{equation}
The $\kappa^u$ and $\kappa^d$ have different signs and we can
conclude that the GPDs $E^u$ and $E^d$ have different signs too.
From (\ref{quarks}) we see that we should have essential
compensation of $E^u$ and $E^d$ effects for $\rho$ meson and
enhancement of their contributions for $\omega$ production.

We calculate  the $A_{UT}$ asymmetry for transversally polarized
protons in \cite{gk08}. The asymmetry is sensitive to
interferences of the amplitudes determined by the $E$ and $H$ GPDs
\begin{equation}\label{aut}
A_{UT}= \propto \frac{\mbox{Im}<E^*>\, <H> }{|<H>|^2}.
\end{equation}
The GPD $H$ was taken from our analysis of the vector meson
electroproduction. Our results for the moments of the $A_{UT}$
asymmetry for the $\rho^0$ production \cite{gk08} describe well
HERMES data \cite{hermesaut}. The predictions  for COMPASS on $t$-
dependence of asymmetry at $W=8 \mbox{GeV}$ shown in Fig.~5  are
in good agreement with preliminary COMPASS data \cite{sandacz}.
Due to essential compensation of the GPD $E$ for valence quark,
the $A_{UT}$ asymmetry for $\rho^0$ is predicted to be quite
small.
\begin{figure}[h!]
\centerline{
\includegraphics[width=0.6\textwidth]{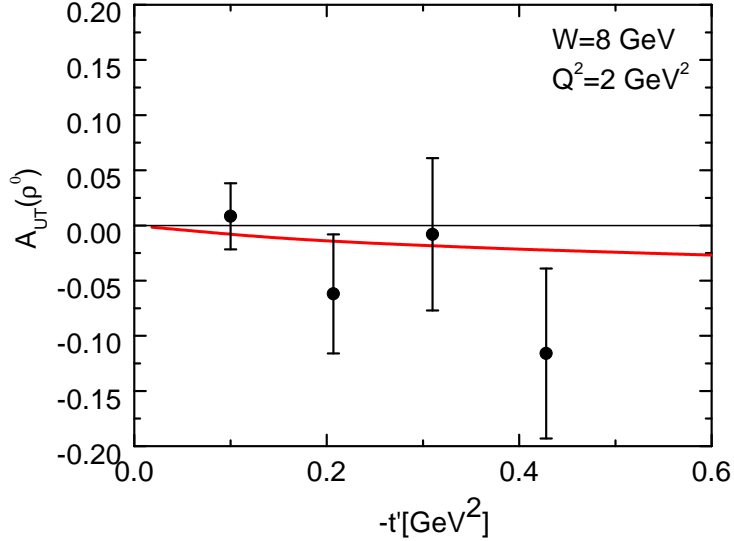}}
\caption{ Predicted $A_{UT}$ asymmetry of the $\rho^0$ production
at COMPASS  with COMPASS data.}
\end{figure}

The predictions for the $A_{UT}$ asymmetry at $W=5 \mbox{GeV}$ and
$W=10 \mbox{GeV}$  were given for the $\omega$,
 $\rho^+$, $K^{*0}$ mesons \cite{gk08}.
It was mentioned that $E^u$ and $E^d$ GPD contributions have the
same sign in the $\omega$ production amplitude, and our results
for $A_{UT}$ asymmetry at HERMES and COMPASS energies are negative
and not small \cite{gk08}.

Predicted $\rho^+$ asymmetry is positive and rather large $\sim$
40\% Fig.~6. However, smallness of the cross section in Fig.~4
does not give a good chance to measure this asymmetry.
\begin{figure}[h!]
\centerline{
\includegraphics[width=0.6\textwidth]{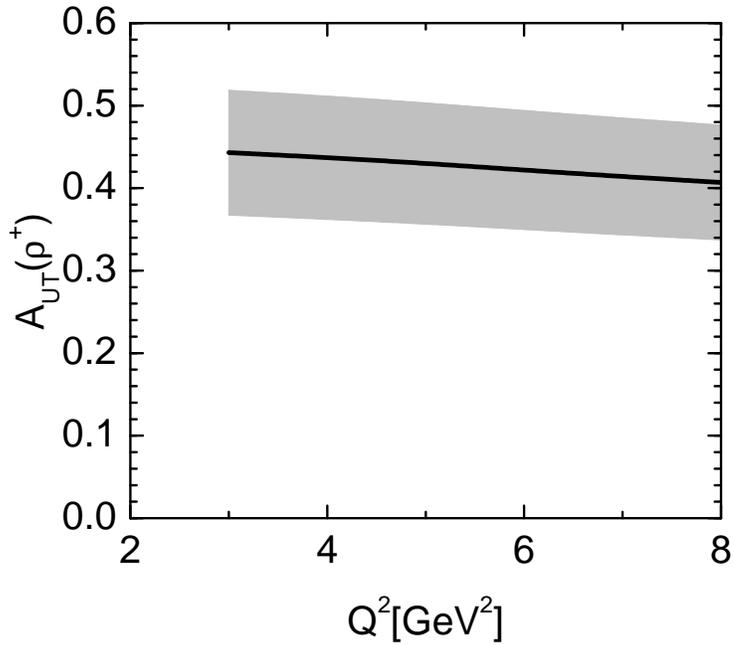}}
\caption{ Predicted $A_{UT}$ asymmetry of the $\rho^+$ production
at HERMES.}
\end{figure}

In our model we found  not small angular momenta for $u$ quarks
and gluons \cite{gk08}
\begin{eqnarray}
<J^u_v>=0.222, \; <J^d_v>=-0.015, <J^g>=0.214,
\end{eqnarray}
which are not far from the lattice results.

\section{Electroproduction of pseudoscalar mesons}
The amplitude of the pseudoscalar meson production with
longitudinally polarized photons ${\cal M}^{P}_{0\nu',0\nu}$
dominates in the process and is determined in terms of polarized
$\tilde H$ GPDs. It can be written at large $Q^2$ \cite{gk09} as
$$
{\cal M}^{P}_{0+,0+} \propto \sqrt{1-\xi^2}\,
                             \,[\langle \tilde{H}^{P}\rangle
  - \frac{2\xi mQ^2}{1-\xi^2}\frac{\rho_P}{t-m_P^2}];\nonumber\\
$$
\begin{eqnarray}\label{pip}
{\cal M}^{P}_{0-,0+} \propto \frac{\sqrt{-t^\prime}}{2m}\,\Big[
\xi \langle \widetilde{E}^{P}\rangle +
2mQ^2\frac{\rho_P}{t-m_P^2}\Big].
\end{eqnarray}
The first terms in (\ref{pip}) represent the handbag contribution
to the pseudoscalar (P) meson production which is calculated
within the MPA with the corresponding transition GPDs in
(\ref{amptt}). For the $\pi^+$ production we have   the $p \to n$
transition GPD where the isovector combination contributes
$\tilde{F}^{(3)}=\tilde{F}^{(u)}-\tilde{F}^{(d)}$.

The second terms in (\ref{pip}) appear for charged meson
production and are connected with the P pole contribution, where
we use the fully experimentally measured electromagnetic form
factor of P meson. The amplitudes with transversely polarized
photons are suppressed as $1/Q$ and can be found in \cite{gk09}.

Using  (\ref{pip}) we calculate all amplitudes with the exception
of ${\cal M}_{0-,++}$ and ${\cal M}_{0+,++}$. It can be shown from
the  angular momentum conservation that we have the following
rule:
\begin{equation}\label{ang}
  M_{\mu'
\nu',\mu \nu} \propto \sqrt{-t'}^{|\mu-\nu-\mu' +\nu'|}
\end{equation}
and the amplitude ${\cal M}_{0-,++}$ should be constant at small
$t'$. However, this amplitude calculated in the handbag approach
behaves as $-t'$ at small momentum transfer. This problem can be
solved if a twist-3 contribution to the amplitude ${\cal
M}_{0-,++}$ is considered. Within the handbag approach this twist-
3 effect can be modeled by the transversity GPDs, in particular
$H_T$, in conjunction with the twist-3 pion wave function
\begin{eqnarray}\label{ht}
{\cal M}^{P,twist-3}_{0-,\mu+} \propto \,
                            \int_{-1}^1 d\xb
   {\cal H}_{0-,\mu+}(\xb,...)\,[H^{P}_T+...O(\xi^2\,E^P_T)].
\end{eqnarray}
For details of calculation of the ${\cal H}_{0-,\mu+}$ amplitudes
see \cite{gk09}.

 The GPD $H_T$ is calculated using
DD form. The transversity PDFs  are connected with $H_T$ as
\begin{equation}
  H^a_T(x,0,0)= \delta^a(x),
\end{equation}
the PDF $\delta$ is parameterized on the basis of the model
\cite{ans}
$$\delta^a(x)=C\,N^a_T\, x^{1/2}\, (1-x)\,[q_a(x)+\Delta
q_a(x)]; $$
\begin{equation}
N^u_T=0.78,\;N^d_T=-1.0.
\end{equation}

It was found that $ H_T$ contributions are essential in the
description of the polarized $\pi$ meson production \cite{gk09}.

We  consider  now the role of the $M_{0+,++}$ amplitude which is
important in some asymmetries and cross section like $\sigma_{T}$,
$\sigma_{TT}$. This amplitude is equal to zero in the leading
twist approximation. The twist-3 contribution to the amplitude
\cite{gk11} has a form similar to (\ref{ht})
\begin{eqnarray}
{\cal M}^{P,twist-3}_{0+,\mu+} \propto \, \frac{\sqrt{-t'}}{4 m}\,
                            \int_{-1}^1 d\xb  {\cal H}_{0-,\mu+}(\xb,...)\; \bar E^{P}_T
\end{eqnarray}
where
\begin{equation}
\bar E^P_T=2\,\tilde H_T+E_T.
\end{equation}

Unfortunately, the information on $E_T$ was obtained only in the
lattice QCD \cite{lat}. It was found that the lower moments of
$E_T$ for $u$ and $d$ quarks are large, have the same sign and a
similar size. This means that we  have an essential compensation
for $\pi^+$ where the combination $E_T^{(3)}=E_T^u-E_T^d$
contributes. The parameters for individual PDFs were taken from
the lattice results \cite{lat} for  $E_T$ moments. The DD model
was used to estimate  $E_T$.

The obtained results on the cross section and  moments of spin
asymmetries \cite{gk09} for the polarized target are in good
agreement with HERMES \cite{hermespi} experimental data.

In Fig. 7, we show the full   cross section of the $\pi^+$
production at HERMES with result for zero $E_T$. The model
describes fine HERMES data, and the $E_T$ contribution is
negligible. The main contribution to the cross section is
determined by the leading-twist longitudinal amplitude.
\begin{figure}[h!]
\centerline{
\includegraphics[width=0.6\textwidth]{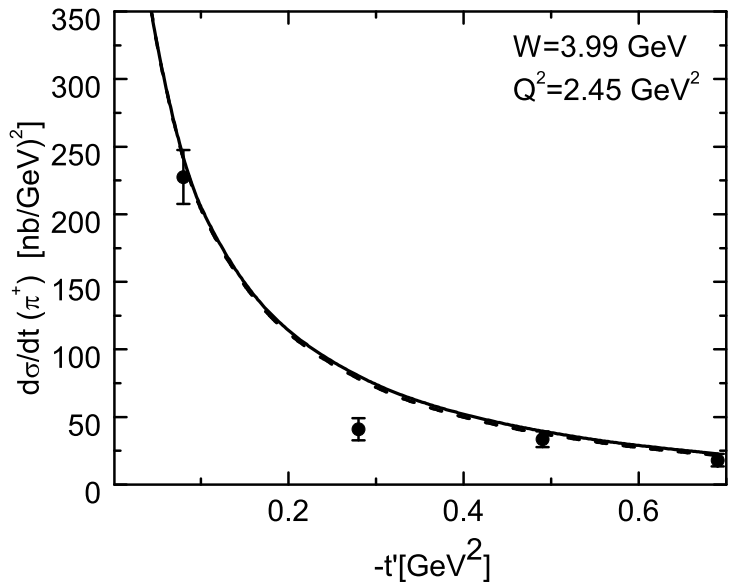}}
\caption{The cross section of the $\pi^+$ production  with HERMES
data. Dashed line -for $E_T=0$}
\end{figure}

The transversity effects are essential in the asymmetry of the
$\pi^+$ production. In Fig.~8, we show our results for the
$\sin{(\phi-\phi_s)}$ moment of $A_{UT}$ asymmetry with and
without the $E_T$ contribution. It can be seen that $E_T$ effects
are very important and improve the description of asymmetry.

\begin{figure}[h!]
\centerline{
\includegraphics[width=0.6\textwidth]{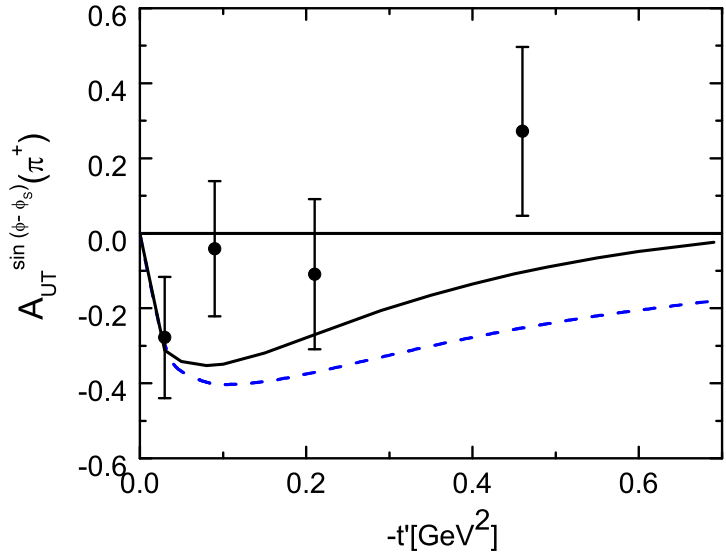}}
\caption{The $\sin{(\phi-\phi_s)}$ moment of $A_{UT}$ asymmetry of
the  $\pi^+$ production  with
 HERMES data. Dashed
line - for $E_T=0$}
\end{figure}

Similar transversity effect is visible in the $A_{UL}$ asymmetry
of the $\pi^+$ production, Fig.~9. Other asymmetries are described
well too in the model \cite{gk09}.

\begin{figure}[h!]
\centerline{
\includegraphics[width=0.6\textwidth]{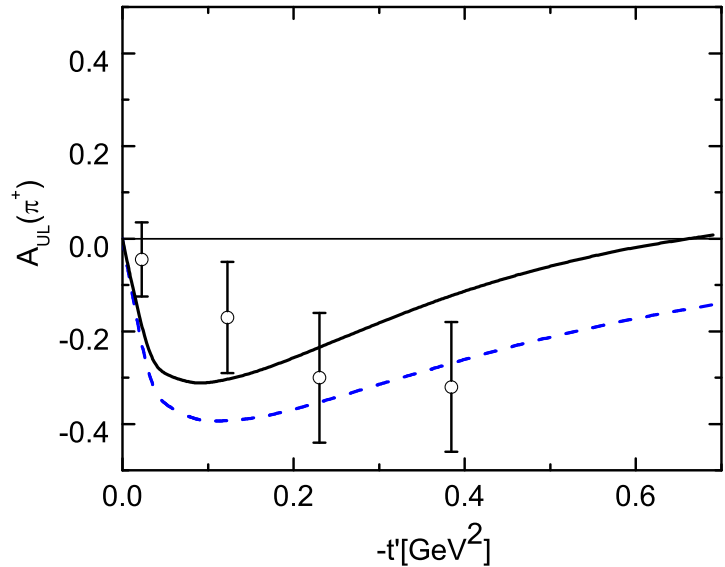}}
\caption{ $A_{UL}$ asymmetry of the $\pi^+$ production  with
 HERMES data. Dashed
line -for $E_T=0$}
\end{figure}

Now we shall discuss out results for the $\pi^0$ production where
a large $E_T$ contribution is expected. Really, for this reaction
we find $E_T^0=2/3\,E_T^u+1/3\,E_T^d$ and we have enhancement of
$E_T$ effects in the $\pi^0$ production instead of compensation
for $\pi^+$. In Fig.~10 we present our results \cite{gk11} for the
cross section of the $\pi^0$ production at HERMES energies.   The
longitudinal cross section, which is determined mainly by the
leading twist contribution and expected to play an essential role,
is much smaller with respect to the $d \sigma_T/dt$ cross section
where the twist-3 $E_T$ contribution is important. Thus, we
observe the predominated role of transversally polarized photons
which are mainly generated by $E_T$. Of cause, this effect will
disappear at large $Q^2$ because twist-3 effects have $1/Q$
suppression with respect to the leading twist contribution.
\begin{figure}[h!]
\centerline{
\includegraphics[width=0.6\textwidth]{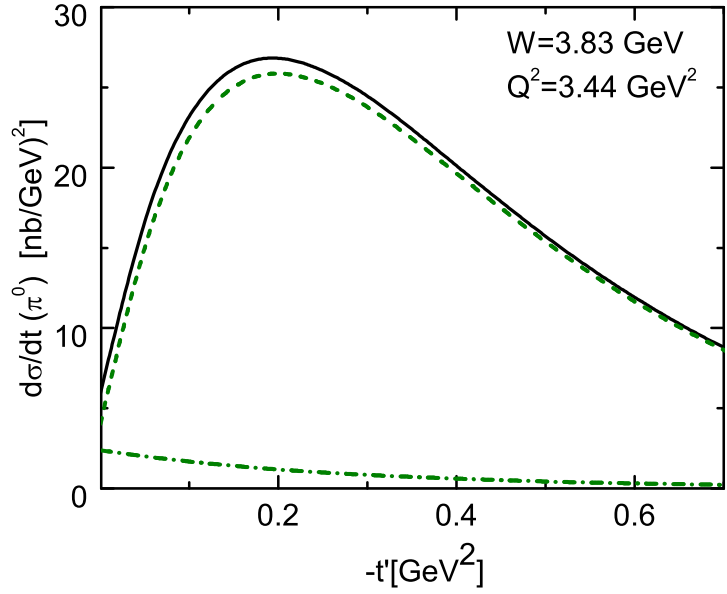}}
\caption{The cross section of the $\pi^0$ production at HERMES.
Full line-full cross section; dashed-dotted- $d \sigma_L/dt$,
dotted line- $d \sigma_T/dt$.}
\end{figure}

In Fig.~11, we show the energy dependence of the $\pi^0$ cross
section at fixed $Q^2$ \cite{gk11}. In the COMPASS energy range
the cross section is not small and can be measured. In Fig.~12,
our results for the $\sin{(\phi-\phi_s)}$ and $\sin{\phi_s}$
moments of $A_{UT}$ asymmetries at HERMES energy are presented
which are predicted to be not small.

\begin{figure}[h!]
\centerline{
\includegraphics[width=0.6\textwidth]{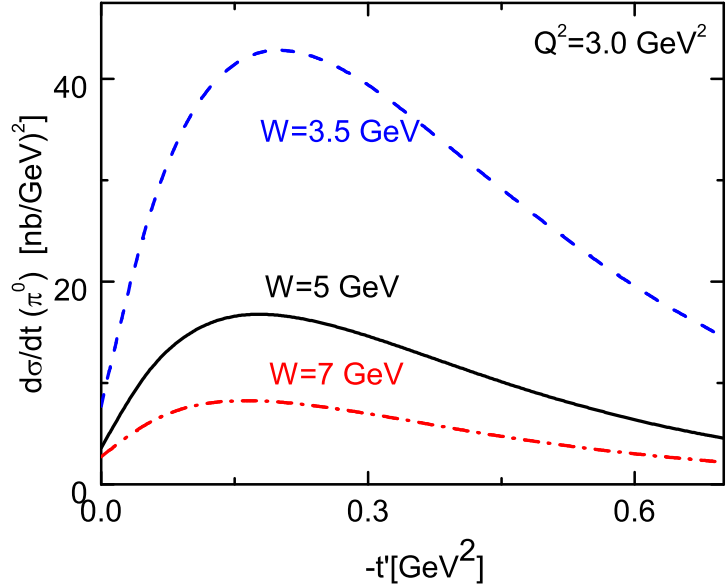}}
\caption{Energy dependence of the cross section of the $\pi^0$
production at $Q^2=3 \mbox{GeV}^2$}
\end{figure}

\begin{figure}[h!]
\centerline{
\includegraphics[width=0.6\textwidth]{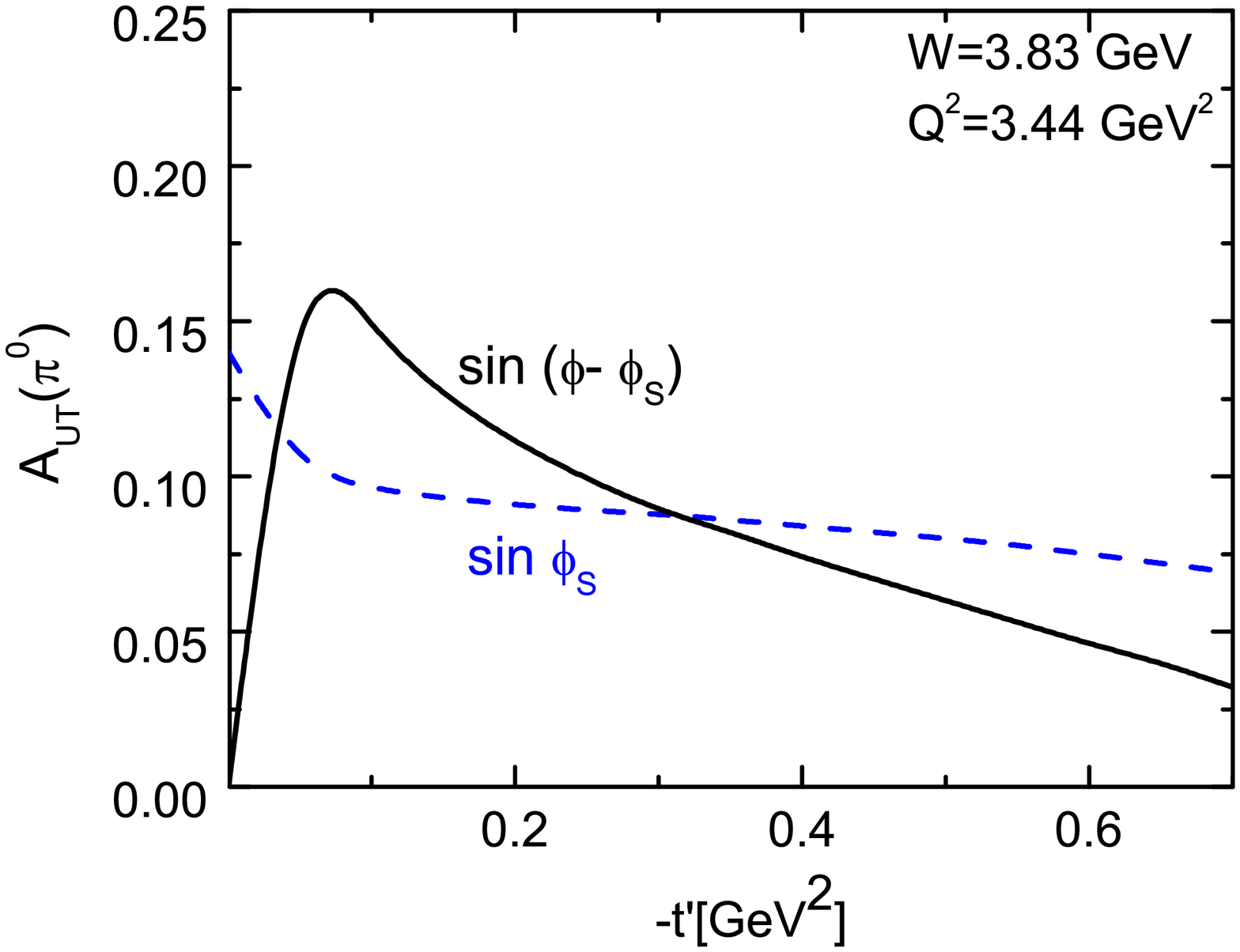}}
\caption{Results for the $\sin{(\phi-\phi_s)}$ and $\sin{\phi_s}$
moments of  the transverse target asymmetries for $\pi^0$
electroproduction }
\end{figure}

 \section{Conclusion and Summary}
In this report, we have studied the light meson electroproduction
within the handbag approach. The  production amplitude at large
$Q^2$ factorizes  into the hard subprocess  and GPDs.  The MPA was
used to calculate the hard subprocess amplitude, where the
transverse quark momenta and the Sudakov factors were considered.
Unfortunately, direct GPDs extraction from observables is
impossible. In our approach we parameterize GPDs using the DD form
and known constraints on PDFs. The obtained results on meson
production were compared with experiment.

It was shown that the $k_\perp^2/Q^2$ corrections in the
propagators of the hard subprocess amplitude were essential in the
description of the cross section at  low $Q^2$. They decrease
$\sigma$ by a factor of about 10 at $Q^2 \sim 3\mbox{GeV}^2$, and
the cross section becomes close to experiment.

The $H$ GPDs calculated using the CTEQ6 parameterization for
gluon, sea and valence quarks  were used to analyze vector meson
production on the unpolarized target. Our results are in good
agreement with the experiment from HERMES to HERA energies.

Information on GPDs $E$ can be obtained from experiments with the
transversely polarized target. These observables at energies $W
\sim 5-10 \mbox{GeV}$ are sensitive to the $H$ and $E$ GPDs for
valence and sea quarks. We model the $E$ GPDs using information on
the Pauli form factors of the nucleon and sum rules. Using these
results we predict the cross sections and $A_{UT}$ asymmetries for
various meson electroproduction \cite{gk08}. The experimental data
are available now only for the $\rho^0$ production and are
described well. We predict not small $A_{UT}$ asymmetry for the
$\omega$ production which most probably may be studied at HERMES
and COMPASS.

The production of pseudoscalar mesons are sensitive to $\tilde H$
and $\tilde E$ GPDs. It was shown that the pion pole and twist -3
effects in the ${\cal M}_{0-,++}$ amplitude determined by $H_T$
contribution are very important in understanding the cross section
and spin asymmetries in the $\pi^+$ production. We estimate
transversity GPD $H_T$ using the model \cite{ans}. The description
of data on the $\pi^+$ production at HERMES is carried out. We can
conclude that information on various GPDs discussed above should
not be far from reality.

We examine the role of transversity GPDs $E_T$ which contribute to
the ${\cal M}_{0+,++}$ amplitude. It was shown that the effects of
$E_T$ GPD are essential in the $\pi^0$ production. At HERMES and
COMPASS energies  the twist-3 $E_T$ effects produce a large
$\sigma_T$ cross section \cite{gk11} which exceeds substantially
the leading twist longitudinal cross section. We predict not small
cross section and spin asymmetries for the $\pi^0$ production. We
expect that this result will be tested in future $\pi^0$
production experiments and shed  light on the role of transversity
effects in these reactions.

We can conclude that the meson electroproduction is a good tool to
probe various  GPDs.

\section{Acknowledgments}
 This work is supported  in part by the Russian Foundation for
Basic Research, Grant  09-02-01149  and by the Heisenberg-Landau
program.

 \bigskip
\nocite{*}
\bibliographystyle{elsarticle-num}
\bibliography{martin}
 
\end{document}